\documentclass[12pt]{iopart}
\usepackage{amssymb}
\usepackage{multirow}
\usepackage{lineno}
\usepackage{graphicx}
\usepackage{bm}
\usepackage{soul}
\usepackage{fancyhdr}
\usepackage[colorlinks=true,citecolor=blue,linkcolor=blue]{hyperref}
\usepackage[usenames]{color}
\usepackage{verbatim}
\usepackage{float}

\begin{document}
	\title[]{Multimillijoule terahertz radiation from laser interactions with microplasma-waveguides}
	\author{Ke Hu$^1$, Longqing Yi$^1$ and
		T\"unde F\"ul\"op$^1$}
	\address{$ˆ1$ Department of Physics, Chalmers University of Technology, 
		41296 Gothenburg, Sweden}
	
	\ead{kehu@chalmers.se}
	
\begin{abstract}
	When a relativistic, femtosecond laser pulse enters a waveguide, the pulse energy is coupled into waveguide optical modes. 
	The longitudinal laser field effectively accelerates electrons along the axis of the channel, while the asymmetric transverse electromagnetic fields tend to expel fast electrons radially outwards. 
	At the exit of the waveguide, the $\sim$${\rm nC}$, $\sim$$10\ {\rm MeV}$ electron bunch converts its energy to a $\sim$$10\ {\rm mJ}$ terahertz (THz) laser pulse through coherent diffraction radiation. 
	In this paper, we present 3D particle-in-cell simulations and theoretical analyses of the aforementioned interaction process. 
	We investigate the process of longitudinal acceleration and radial expulsion of fast electrons, as well as the dependence of the properties of the resulting THz radiation on laser and plasma parameters and the effects of the preplasma.
	The simulation results indicate that the conversion efficiency of energy can be over $5\%$ if the waveguide length is optimal and a high contrast pump laser is used. These results guide the design of more intense and powerful THz sources.
	\\
	\\
	Keywords: multimillijoule terahertz radiation, microplasma waveguide, electron acceleration, laser-plasma interaction
\end{abstract}

\maketitle

\section{Introduction}

During the last decade, terahertz (THz) sources based on laser-plasma interactions have attracted considerable attention due to their potential to produce ${\rm GV/cm}$, ${\rm mJ}$ level THz radiation. 
Such powerful THz sources would open up new regimes to investigate/manipulate physical systems across a broad range of research areas, ranging from biology to astrophysics \cite{Hebling2008,Pickwell2006,Siegel2004,Globus2003}.
Laser-driven underdense plasmas usually deliver a few-microjoule THz radiation in experiments and the THz energy saturates with increasing pump laser intensity \cite{Leemans2003,Clerici2013,Xie2006,Dechard2018}. However, particle-in-cell (PIC) simulations have predicted that $\sim$$1\ {\rm GV/cm}$, $\sim$$1\ {\rm mJ}$ THz pulses can be
obtained by adopting a tailored pump laser \cite{Chen2016} or by obliquely irradiating a underdense plasma slab with sub-100-$\mu$m thickness \cite{Wu2008}. 
Employing laser-solid interactions appear even more promising \cite{Liao2016,Ding2016,Liao20191}.
As a solid foil is irradiated by a pump laser, coherent transition radiation in the THz range is emitted in the backward and forward directions, when the accelerated electrons pass through the front and rear surfaces
of the target, respectively. Forward THz pulses with energy over $10\ {\rm mJ}$ have been experimentally demonstrated by Liao et al \cite{Liao20192}, with conversion efficiency at the level of $0.1\%$.

The main factor in shaping the properties of coherent transition radiation from laser-driven solid targets is the quality of the electron beam \cite{Schroeder2004,Ding2019}. The radiation is coherent if the bunch length is shorter than the radiated wavelength of interest, and the THz energy improves strongly with decreasing beam divergence. In the coherent regime, the radiated energy is proportional to the square of beam charge. Therefore, an electron beam with high charge, high energy and small divergence is required for generating powerful THz radiation. For solid foil targets, the electron beam usually suffers from large divergence, and this prevents scaling the scheme towards higher THz energies \cite{Liao20162}.

In order to acquire high-quality electron beams and high-energy THz radiation, schemes utilizing micro-structured targets have been proposed. 
3D PIC simulations demonstrate strong electron emission at the micro-scaled target edge, which leads to a THz energy of over $10\ {\rm mJ}$ with a $1$-${\rm J}$ pump laser \cite{Hu2020}. 
Another study employs solid foil targets covered with aligned nanorod arrays in experiments;  the resulting efficiency is enhanced by an order of magnitude compared to a solid foil target \cite{Mondal2017}.

Among these structured targets is the microplasma waveguide (MPW); not only does it  suppress the transverse diffraction of the pump laser, but it also enhances the longitudinal acceleration field \cite{Gong2019,Shen1991}. 
Such targets have already shown their potential in electron acceleration \cite{Snyder2019,Xiao2016}, X-ray generation \cite{Yi20161,Yi20162}, production of ion beams \cite{Zou2015,Kluge2012} and manipulation of relativistic laser pulses \cite{Ji20161}. 
In our previous work, simulations show that high charge ($\sim10\ {\rm nC}$), high energy ($\sim100\ {\rm MeV}$) and well-collimated ($10^{\circ}$) electron bunches can be produced and accelerated by the transverse magnetic modes \cite{Yi2019}. Their energies are converted to strong THz emission through coherent diffraction radiation (CDR) when they exit the MPW. 
Although that study has shown that an efficiency over $1\%$ can be realized, even more powerful THz output can be reached by target optimization. 
The goal of the present paper is to optimize the laser and target parameters for highly efficient THz generation. This will be done by investigating the  dynamics of fast electrons inside a MPW, in particular the effects of several target parameters, including target length and preplasma scale length, by means of 3D PIC simulations and  analytic theory.  
It is found that the efficiency can be over $5\%$ when the MPW length is optimal and the preplasma scale length is small.

The paper is organized as follows: Sec.~II introduces the optical modes and electron dynamics inside a MPW. Sec.~III discusses properties of THz radiation as well as its dependence on laser-plasma parameters. 
In Sec.~IV, the effect of preplasma is studied.
At last, a brief summary is given in Sec.~V. 

\section{Electron dynamics}
The setup of a laser-MPW THz source is shown in Fig.~1(a). A  linearly polarized laser pulse is tightly focused onto the entrance of a cylindrical waveguide along the $x$ axis, from the left. 
The electrons get accelerated inside the MPW and finally convert their energy to a relativistic half-cycle THz radiation pulse when leaving the channel.
The interaction process is explored using the 3D PIC code EPOCH \cite{arber}.
The dimensions of the simulation box are $x \times y \times z=80\ {\rm \mu m} \times 100\ {\rm \mu m} \times 100\ {\rm \mu m}$ with grid steps $dx \times dy \times dz=0.05\ {\rm \mu m} \times 0.1\ {\rm \mu m} \times 0.1\ {\rm \mu m}$. 
The pump laser pulse has a temporal FWHM duration of $T_0=35\ {\rm fs}$ and normalized amplitude of $a_0=eE_0/mc\omega_0=10$, where 
$c$ is the speed of light, $m$ is the electron mass, $e$ is the unit charge, $\lambda_0=1\ {\rm \mu m}$ is the laser wavelength and $\omega_0=2\pi c/\lambda_0$ is the angular frequency. 
The spot size of the pulse is $w_0=3\ {\rm \mu m}$.
The MPW target has a length of $L=20\ {\rm \mu m}$ and a density of $n_0=15n_c$, where $n_c=\epsilon_0m\omega_0^2/e^2$ is the critical density. 
Its wall thickness and inner radius are $5\ {\rm \mu m}$ and $r_0=3\ {\rm \mu m}$, respectively. The entrance of the MPW is placed at $x=1\ {\rm \mu m}$. 

Figure~1(a) presents the $40.2$-${\rm mJ}$ THz radiation pulse shown with rainbow color scale. 
The electric fields are considered in spherical coordinates with the origin at the exit of the MPW. 
The polar component $E_{\theta}$ contains $97\%$ of the THz energy, since the coherent diffraction radiation is intrinsically radially polarized. 
The peak electric amplitude reaches $7\ {\rm GV/cm}$, indicating that the THz pulse is relativistic. The optical-to-THz conversion  efficiency  is $6.7\%$, much higher than can be achieved by any other state-of-art laser-plasma THz source. 	

\begin{figure}[t]
	\centering
	\includegraphics[width=0.48\textwidth]{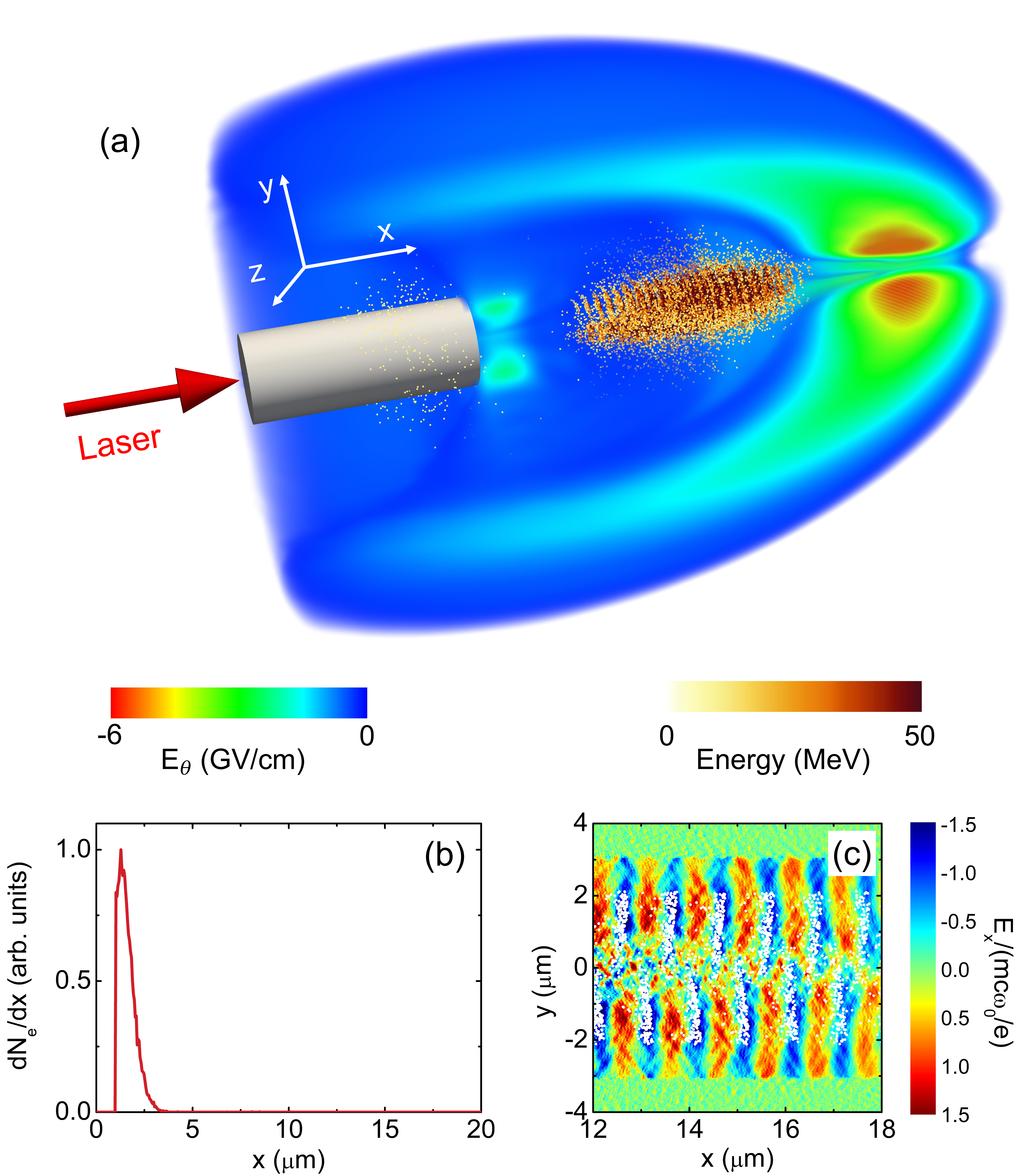} 
	\caption{
		(a) 3D schematic setup of the THz radiation source based on an MPW. 
		The orange dots are fast electrons (with relativistic gamma factor $\gamma>10$) at simulation time $t=200\ {\rm fs}$ and the color represents their energies.
		The polar component of the THz field $E_{\theta}$ (frequency $f<60\ {\rm THz}$) at $t=267\ {\rm fs}$ is shown with rainbow colorscale, where a quarter is removed to display the intensity inside. 
		(b) The distribution of the initial longitudinal positions of the  electrons shown in (a).
		(c) Cross section of the longitudinal electric field inside the MPW
		at $t=107\ {\rm fs}$ in the $x-y$ plane at where $z=0\ {\rm \mu m}$.
		White dots represent projections of the positions of fast electrons on the $x-y$ plane.
	} 		
\end{figure}

Strong THz radiation is emitted when the energetic electron bunch, represented by the orange dots in Fig.~1(a), leaves the channel.
We track those electrons and plot the distribution of their initial locations along the $x$ direction in Fig.~1(b).
We note that almost all these electrons originate from the vicinity of the entrance of MPW. This can be attributed to the strong diffracted laser light generated during the violent impact of the laser and onto the MPW front surface. 
Such diffraction-induced injection is the dominant mechanism for electron injection as long as $r_0 \leq w_0$. 
In the next section we will show that the THz radiation under this condition is much more powerful than that in the case of $r_0 > w_0$. 
Since our main interest is high-energy THz radiation, in the following,  we focus on the interaction when $r_0 \leq w_0$ and  assume all fast electrons are to be injected at the MPW entrance.

We then proceed to discussing the propagation of the pump laser  inside a MPW.
Normally, many optical modes are excited simultaneously, and their intensities mainly depend on the waveguide radius.
For the micro-scale waveguide used in our scheme, most of the laser energy is coupled into the fundamental waveguide mode and higher modes can be ignored, which effectively simplifies our analysis.
Considering the shape of the target, it is appropriate to express the electromagnetic fields in cylindrical coordinates ($x$, $r$, $\phi$) \cite{Yi20162}:

\begin{eqnarray}
E_{x}=E_0\frac{k_t}{k}J_1(k_t r){\rm sin}(\phi){\rm cos}(\Phi),\\
B_{x}=-B_0\frac{k_t}{k}J_1(k_t r){\rm cos}(\phi){\rm cos}(\Phi),\\
E_r=-\frac{E_0}{2}\left[\frac{k+k_x}{k_t}J_0\left(k_t r\right)  + \frac{k-k_x}{k_t}J_2\left(k_t r\right) \right]{\rm sin}(\phi){\rm sin}(\Phi),\\
B_r=\frac{B_0}{2}\left[\frac{k+k_x}{k_t}J_0\left(k_t r\right)  - \frac{k-k_x}{k_t}J_2\left(k_t r\right) \right]{\rm cos}(\phi){\rm sin}(\Phi),\\
E_{\phi}=-\frac{E_0}{2}\left[\frac{k+k_x}{k_t}J_0\left(k_t r\right)  + \frac{k-k_x}{k_t}J_2\left(k_t r\right) \right]{\rm cos}(\phi){\rm sin}(\Phi),\\
B_{\phi}=-\frac{B_0}{2}\left[\frac{k+k_x}{k_t}J_0\left(k_t r\right)  - \frac{k-k_x}{k_t}J_2\left(k_t r\right) \right]{\rm sin}(\phi){\rm sin}(\Phi).
\end{eqnarray}

where $E_0$ and $B_0$ are the amplitude of the electric and magnetic components of the pump laser, respectively. $k=2\pi/\lambda_0$ is the wave number in vacuum and $\Phi$ is the phase within the laser pulse. 
$J_\alpha(x)$ denotes the Bessel function of the first kind of order $\alpha$.
$k_x$ and $k_t$ are the longitudinal and transverse components of the wave number inside the MPW, respectively, which satisfy
$k_t={x_1}/{r_0}$ and $k_t^2+k_x^2=k^2$.
$x_1$ is a parameter obtained by numerically solving an eigenvalue equation \cite{Shen1991}. For the interaction geometry defined here, $x_1=2.5$ can be considered as a constant \cite{Yi20162}. 

Structures of the longitudinal electric field $E_x$ in the $x-y$ plane are presented in Fig.~1(c). The amplitude of the acceleration gradient is $4\ {\rm TV/m}$, much higher than the longitudinal field of the pump laser in vacuum. 
Even though fast electrons acquire most of their energy from longitudinal acceleration, we note that only a small amount of laser energy is coupled to the longitudinal fields, namely $E_{x}^2 \ll (E_r^2+E_{\phi}^2)$. 

One important feature of $E_x$ is that its phase velocity is superluminal, which means that energetic electrons undergo first an acceleration stage when they are in desirable phases (see Fig.~1(c)), then a deceleration stage after the dephasing effect takes place. The phase varies from $\Phi=0$ to $\Phi=\pi$ at the acceleration stage.
The maximum acceleration distance $L_{acc}$ satisfies \cite{Yi2019}

\begin{equation}
L_{acc} \propto r_0^2.
\end{equation}
From Eq.~(1), we can deduce that the peak acceleration gradient is proportional to $a_0/r_0$. Therefore the maximum relativistic gamma factor fulfills
\begin{eqnarray}
\gamma_{max} \propto r_0a_0.
\end{eqnarray}

\begin{figure}[t]
	\centering
	\includegraphics[width=0.48\textwidth]{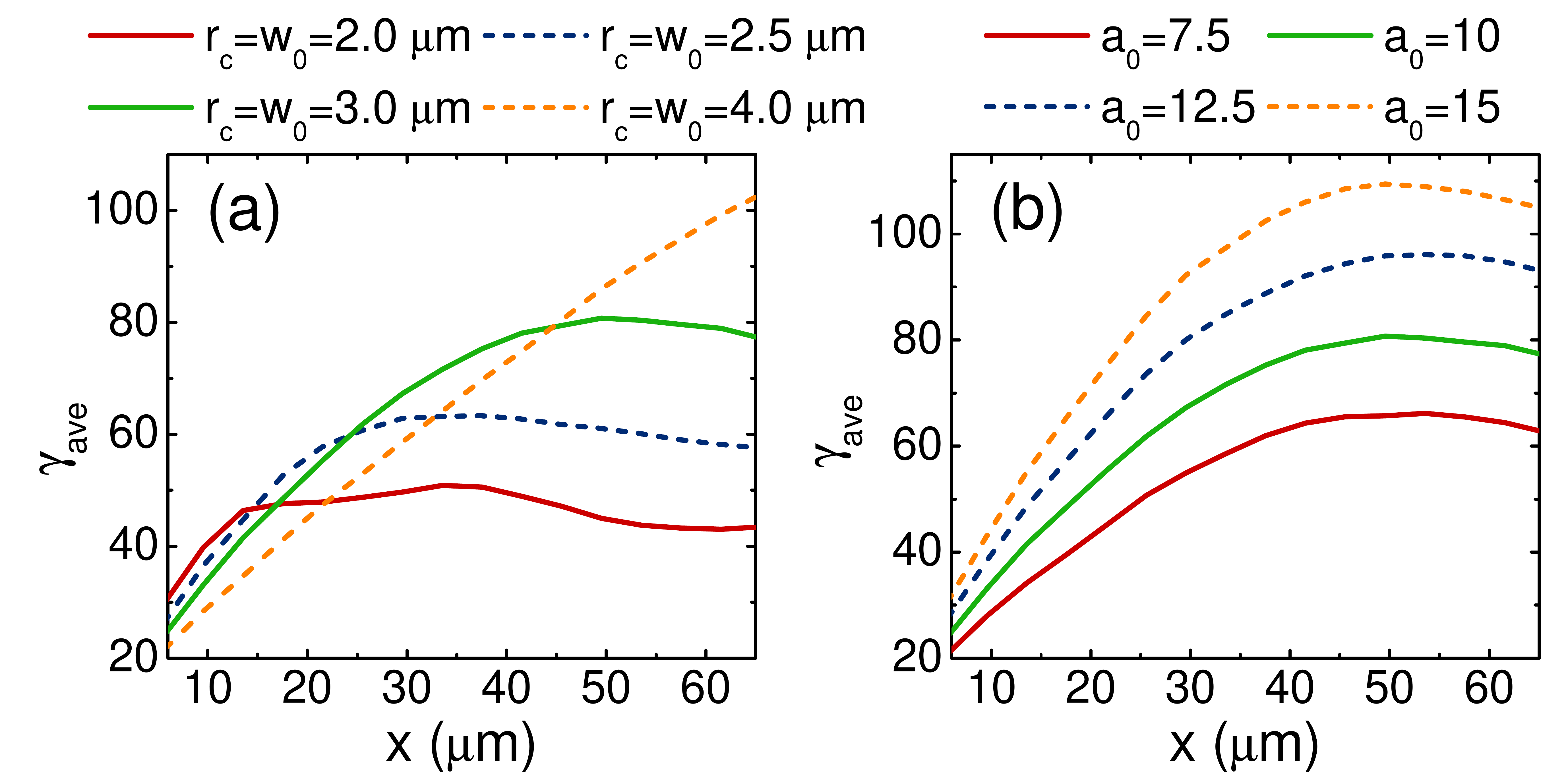} 
	\caption{
		Average $\gamma$ of internal fast electrons plotted against propagation distance of the pump laser for different $r_0$ (a) and $a_0$ (b). $a_0=10$ is fixed in (a), channel inner radius $r_c=3\ {\rm \mu m}$ is fixed in (b) and laser spot size $w_0=r_c$ is satisfied in both panels. 
	} 		
\end{figure} 

Next, we show a series of simulations, in which the laser waist size equals  the channel inner radius, ranging from $2.0\ {\rm \mu m}$ to $4.0\ {\rm \mu m}$. A few-cycle pump laser ($T_0=10\ {\rm fs}$) is adopted in order to lower the length of the electron bunch, making it easier to distinguish the acceleration or deceleration stage. 
As soon as the laser enters the waveguide, excited electrons ($\gamma>3$) are selected out and their dynamics are tracked every four laser cycles during the simulation. At each moment we exclude escaping electrons ($r>r_0$), and only consider energetic electrons ($\gamma>10$) that are inside the tube, which we call internal fast electrons (IFEs) hereafter.

The average gamma factor of IFEs versus propagation distance of the pump laser, for different $w_0$ and $r_0$ are plotted in Fig.~2(a). At the acceleration stage, the average electron energy increases rapidly for small $r_0$, since the acceleration field is inversely proportional to MPW radius.
However, the violent acceleration phase only lasts for a short time and the saturation energy is below $25\ {\rm MeV}$ when $r_0=2\ {\rm \mu m}$.
On the contrary, when the channel radius is larger, the acceleration is weaker but lasts for a longer time, finally leads to a higher saturation energy.

\begin{figure}[t]
	\centering
	\includegraphics[width=0.48\textwidth]{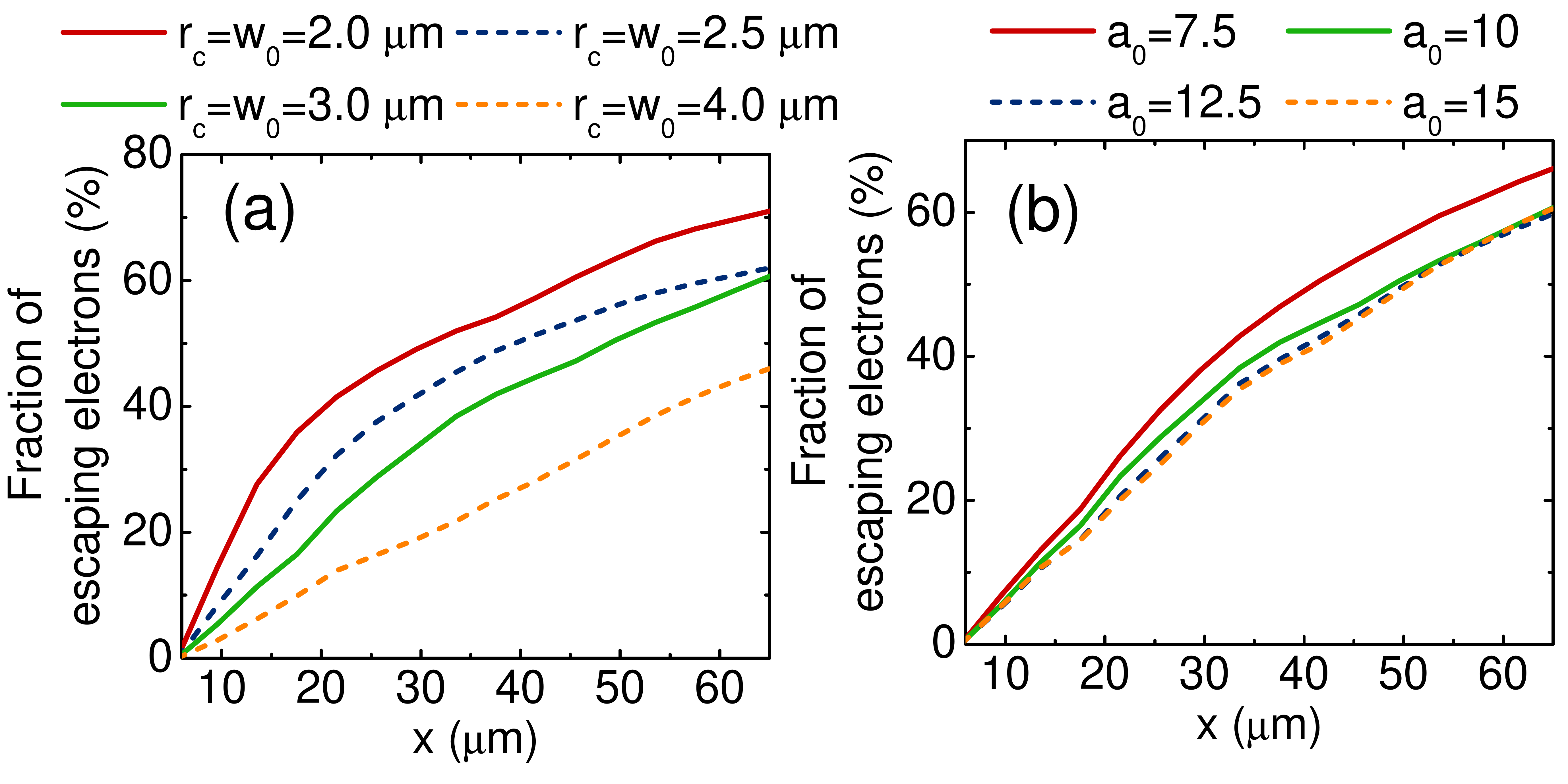} 
	\caption{Fraction of radially escaping electrons
		plotted against propagation distance of the pump laser for different $r_0$ (a) and $a_0$ (b).
		$a_0=10$ is fixed in (a), $r_c=3\ {\rm \mu m}$ is fixed in (b), and $w_0=r_c$ is satisfied in all panels.
	} 		
\end{figure} 

Similarly, in Fig.~2(b), we present the effect of laser amplitude on the acceleration process, in the case of a fixed channel radius $r_0=3\ {\rm \mu m}$. 
The electron energies reach their maximum values at almost the same time, since the acceleration length is not related to the pump laser amplitude (Eq.~(7)).
Furthermore,  the saturation energy increases with laser amplitude, in agreement with Eq.~(8).

We now proceed to discuss the effect of transverse optical modes on electron dynamics. Using Eqs.~(3)-(6), the transverse force acting on the IFEs can be derived as
\begin{eqnarray}
F_{\bot}=e\sqrt{(E_r-B_{\phi})^2+(E_{\phi}-B_r)^2}\propto \frac{a_0}{r_0^2} {\rm sin}\left(\Phi \right).
\end{eqnarray} 
The force can lead to a maximum transverse displacement $r_{max}$, that scales as
\begin{eqnarray}
r_{max} \propto a_0r_c^2.
\end{eqnarray}

At the same time, the relative phase varies from $\Phi=0$ to $\Phi=\pi/2$, which means that the IFEs have passed approximately half of the maximum acceleration distance in the $x$ direction. 
For small channel radius, the maximum acceleration distance is short, hence a large fraction of fast electrons tend to be scattered and escape earlier. Those escaping electrons transmit into the target bulk through inner boundaries of the MPW before reaching the channel exit, thus making a negligible contribution to THz emission.

Figures 3(a) and (b) show the fraction of escaping electrons among the energetic ones versus the pump laser's propagation distance for different channel radii and laser amplitudes, respectively. 
Electrons escape from the channel fast when the channel radius is small, same as what we predicted above.
Since the maximum acceleration distance is not related to the laser amplitude, the rates of reduction in electron charge are approximately equal for different $a_0$. 

\begin{figure}[b]
	\centering
	\includegraphics[width=0.48\textwidth]{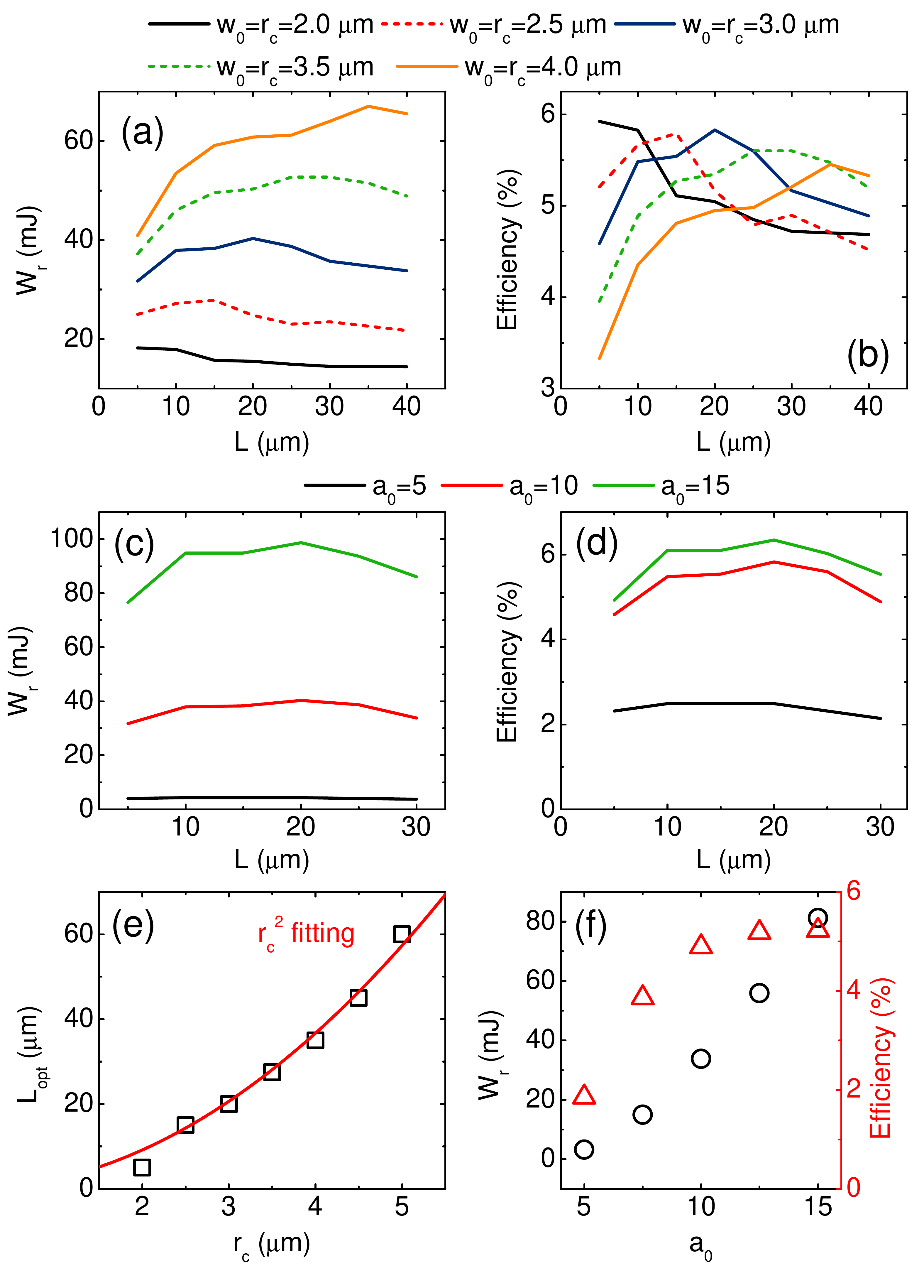} 
	\caption{
		Dependence of THz energy ((a) and (c)) and conversion efficiency of energy ((b) and (d)) on the length of MPW for different $r_0$ ((a) and (b)) and $a_0$ ((c) and (d)). 
		$a_0=10$ is fixed in (a) and (c), while $r_c=w_0=3\ {\rm \mu m}$ is fixed in (b) and (d).
		(e) The optimal length of MPWs vs $r_c$ for fixed $a_0=10$. 
		(f) Peak THz radiation energy and conversion efficiency vs varying laser amplitude $a_0$ for fixed $r_c=3\ {\rm \mu m}$. $w_0=r_c$ is satisfied in (e) and (f).
	} 		
\end{figure} 

\section{Terahertz radiation}

The radiated THz energy scales as the square of the beam charge that leaves the waveguide.
The electron beam consists of electrons that fulfill the following two conditions:
(i) have high enough energies to escape through the exit;
(ii) are not expelled from the channel radially before arriving in the exit.
A short MPW is not favorable for the pump laser to deposit its energy to more IFEs, whereas in a long target, more fast electrons will escape during the long propagation.
Therefore the MPW has an optimal length $L_{opt}$
at which the number and energy of IFEs are most advantageous to the emission of THz radiation.
The scattering angle of the IFEs can be roughly estimated by ${\rm tan}(\Theta) \approx r_{max}/L_{acc}$.
Therefore, using Eq.~(7) and Eq.~(10), we obtain
\begin{eqnarray}
L_{opt}=\frac{r_{0}}{{\rm tan}(\Theta)} \propto r_0^2.
\end{eqnarray}

\begin{figure}[t]
	\centering
	\includegraphics[width=0.3\textwidth]{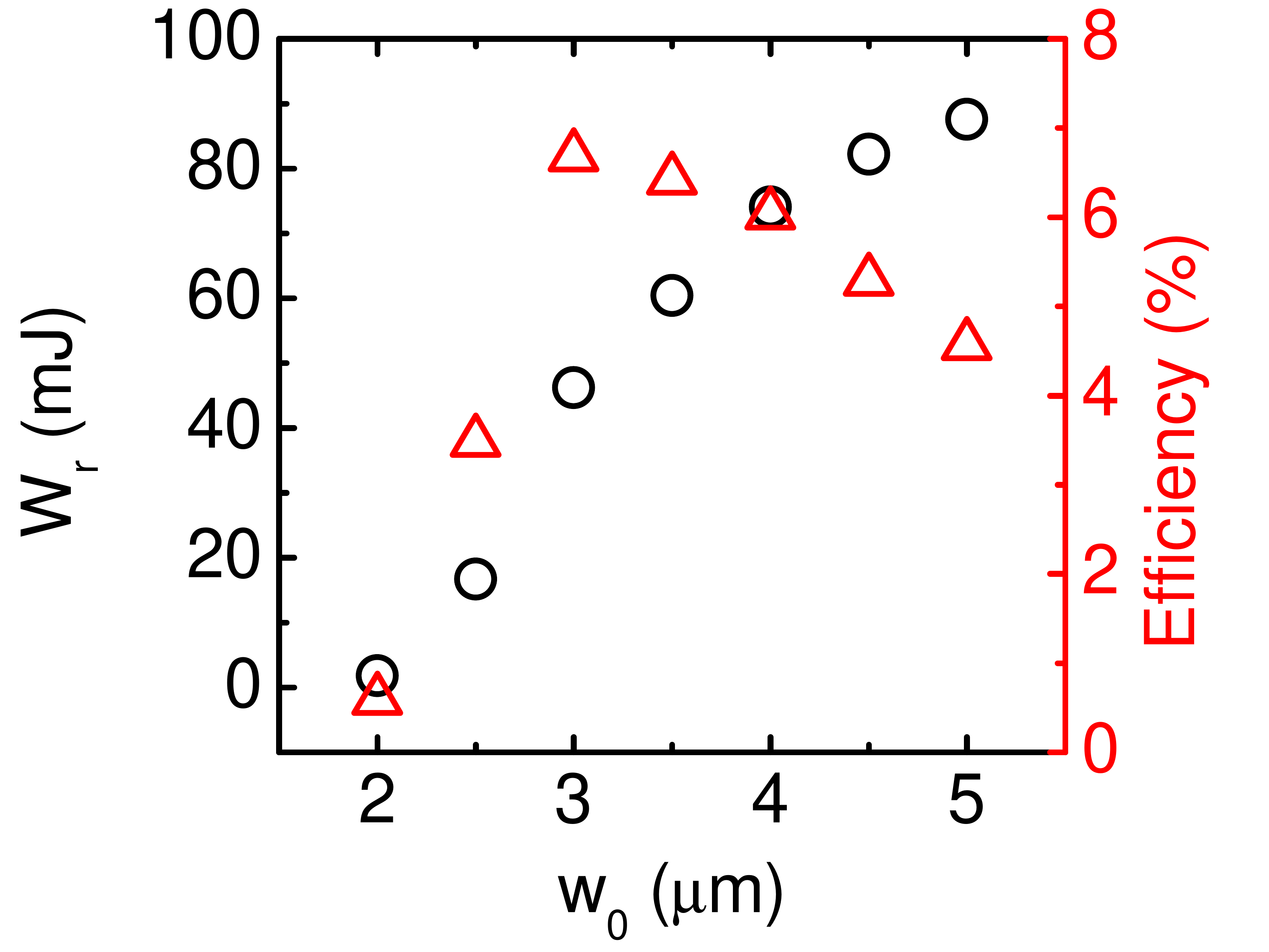} 
	\caption{
		THz radiation energy and conversion efficiency as function of laser spot size $w_0$. The other parameters are $r_c=3\ {\rm \mu m}$, $a_0=10$ and $L=20\ {\rm \mu m}$.
	} 		
\end{figure} 
A series of 3D PIC simulations are conducted to investigate the optimal MPW length.  The laser pulse duration is set to $T_0=35\ {\rm fs}$. 
Since the THz frequency depends on the length of the electron bunch and it is close to the pump laser duration, $T_0=35\ {\rm fs}$ leads to a central THz frequency of about $3\ {\rm THz}$.
We  introduce preplasma at the inner surface of the MPW, in order to better simulate the interaction in realistic conditions. The preplasma has a density profile $n(r)=n_0{\rm exp}\left[ -(r-r_0)^2/\sigma_0^2 \right]$, where $\sigma_0$ is the scale length.
This leads to an effective channel radius $r_c$, which satisfies 
$n(r_c)=n_c$.
In the previous equations, $r_0$ should be replaced with $r_c$ in the presence of preplasma.
Here, the scale length is set as $\sigma_0=0.2\ {\rm \mu m}$; a discussion on its effect can be found in the next section. 

Figures 4(a) and 4(b) show the THz energy and conversion efficiency against the length of MPW for different MPW radius $r_c$, respectively. 
It can be observed that the energy of THz radiation is typically over $10\ {\rm mJ}$. 
As the MPW length is increased, the THz energy first increases then reduces in every case except for $r_c=2.0\ {\rm \mu m}$. 
The reason is that when the tube radius is small, the effect of expulsion is so severe that it has more significant impact than the effect of acceleration in a very early stage.
Figure 4(c) shows the relationship between optimal MPW length and channel radius.
The dependence of $L_{opt}$ on $r_c$ has a squared scaling as predicted by Eq.~(11).

Similarly, in Figs.~4(c) and 4(d), we present the THz energy and conversion efficiency under different laser amplitudes. When the laser spot size and channel radius are fixed at $w_0=r_c=3\ {\rm \mu m}$, the optimal MPW length is the same for laser amplitudes ranging from $a_0=5$ to $15$. From Fig.~4(c), we see that the peak THz energy reaches $99\ {\rm mJ}$ in the case of $a_0=15$ (corresponding to a pump laser energy of $1.6\ {\rm J}$). 
In addition, Fig.~4(f) illustrates that the peak efficiency increases with $a_0$ and finally saturates at about $6\%$ when $a_0>10$. Keeping in mind that the energy spectrum of IFEs satisfies an approximately Maxwellian distribution, a small laser amplitude leads to a small saturation energy, thus many relatively low-energy electrons (distributed on the low-energy part of the spectrum) are not energetic enough to escape from the MPW. This accounts for the rapidly dropping efficiency as $a_0$ is decreased below $10$. 

Finally, Figure 5 shows a scan of the THz energy and efficiency plotted against laser spot size $w_0$. $r_c=3\ {\rm \mu m}$ remains unchanged to ensure the same optimal MPW length $L=20\ {\rm \mu m}$ for all cases. 
The THz energy increases with laser spot size, while the efficiency reaches its maximum at $w_0=r_c$. 
Increasing $w_0$ leads to enhanced electron excitation at the entrance and more powerful electron acceleration inside the channel, thus helps for the boost of  THz energy. 
However, for $w_0>r_c$, increasing $w_0$ means a smaller fraction of the laser that can make its way to the channel, which is responsible for the decrease of efficiency.


\begin{figure}[t]
	\centering
	\includegraphics[width=0.48\textwidth]{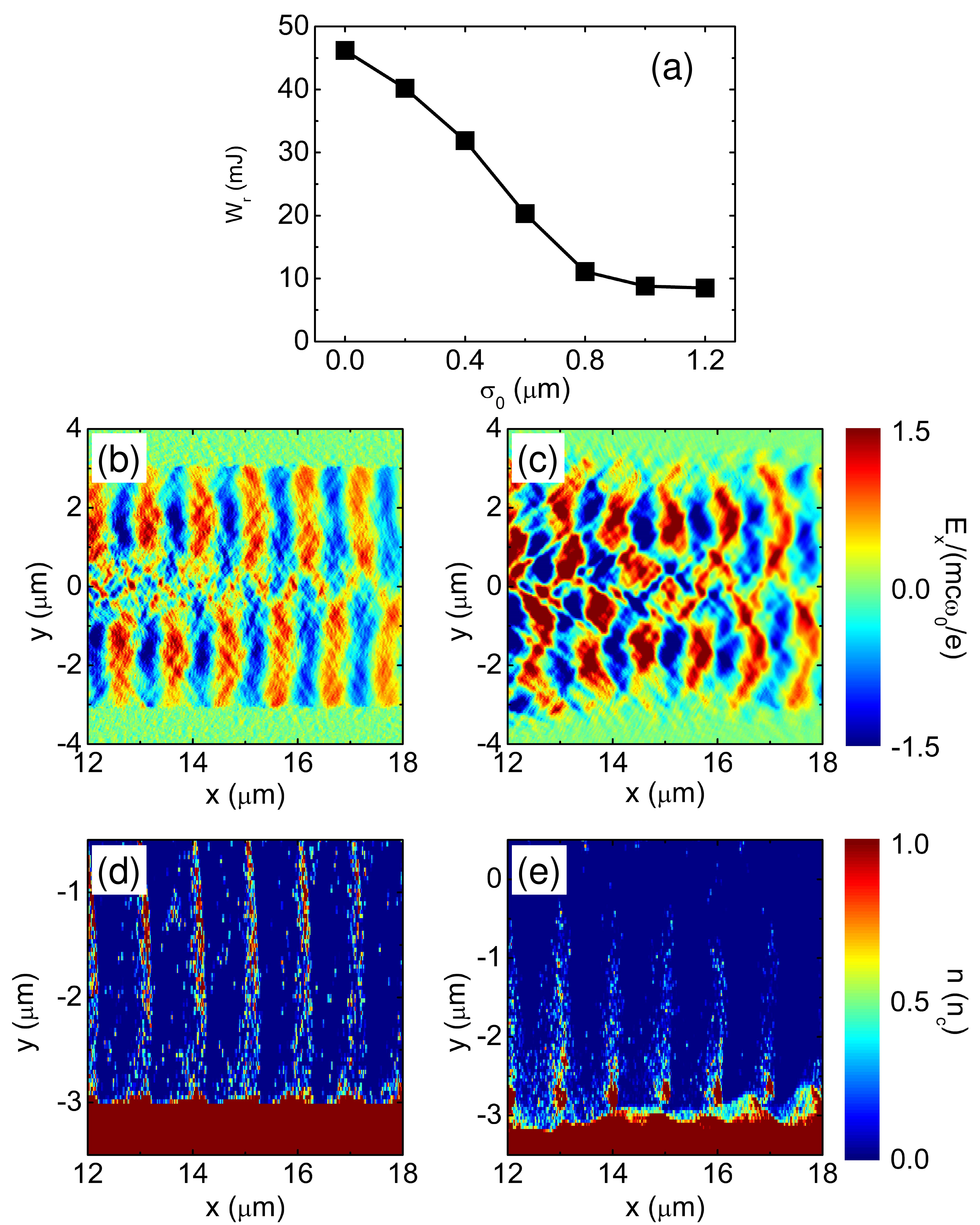} \\ 
	\caption{
		(a) THz radiation energy as a function of the preplasma density scale length.
		(b)(c) show the distribution of longitudinal electric field inside the MPW at $t=200\ {\rm fs}$ for $\sigma_0=0\ {\rm \mu m}$ and $\sigma_0=0.8\ {\rm \mu m}$, respectively.
		(d)(e) present the density distribution of inner boundaries of MPWs under the same condition as in (b) and (c), respectively.
	} 		
\end{figure} 

\section{Preplasma effects}
In relativistic laser interaction with micro-scale channel target,  the preplasma condition, in particular the scale length of the density profile on the inner boundary, has a significant impact on the number and energy of fast electrons, as well as the energy of THz radiation.
In Fig.~5(a), THz energy is plotted against the scale length $\sigma_0$ in the case of $a_0=10$, $L=20\ {\rm \mu m}$, $r_c=w_0=3\ {\rm \mu m}$. 
We note that the THz energy decreases with increasing preplasma scale length. When $\sigma_0=0.8\ {\rm \mu m}$, the THz radiation has an energy of $18\ {\rm mJ}$, corresponding to about $25\%$ of the case without preplasma.

The main reason for the detrimental effect of the preplasma is that the distribution of electromagnetic fields inside a MPW is modified. To  illustrate this, we use 3D PIC simulations with higher resolution ($dx \times dy \times dz =0.02\ {\rm \mu m} \times 0.04\ {\rm \mu m} \times 0.04\ {\rm \mu m}$), while the other parameters remain the same. 
Figures 5(b) and 5(c) present the distribution of acceleration gradient $E_x$ for $\sigma_0=0$ and $\sigma_0=0.8\ {\rm \mu m}$, respectively. The snapshot was taken at the moment when the pump laser pulse has propagated a distance of
\textbf{$20\ {\rm \mu m}$}. It is evident that the electric field in Fig.~5(c) is greatly distorted and is no longer radially uniform, resulting in less IFEs that can be accelerated steadily for a sufficiently long time.

In essence, the distortion of optical modes can be attributed to the oscillation of inner waveguide surfaces, induced directly by the radial component of the laser field. Such oscillations are violent in the region of near-critical-density preplasmas \cite{Lichters1996,Bulanov1994}, which can be illustrated by a snapshot of density profile around the inner plasma surface when $\sigma_0=0.8\ {\rm \mu m}$, plotted in Fig.~6(e). The amplitude of oscillation is so large that overdense electron clusters are extracted from the surface and then reinjected into the bulk plasma. In contrast, in Fig.~6(d), one sees mild density oscillations in the case of a step boundary of overdense plasma. Therefore, in order to obtain high THz energy, it is required to lower the scale length of preplasmas by using high contrast pump lasers.

\section{Conclusion}
We have studied a mechanism of relativistic terahertz radiation generation based on laser interaction with micro-plasma-waveguides via coherent diffraction radiation. The critical underlying physical processes involve acceleration and expulsion of fast electrons. 
The former can be attributed to powerful longitudinal component of the optical modes inside the MPW, while the latter is induced by a weak transverse force caused by the asymmetry in the transverse components of these optical modes. 
The two processes for different laser amplitudes and waveguide radii are investigated via a theoretical model and 3D PIC simulations, in order to find parameters that optimize the optical-to-THz conversion efficiency. 

It is found that the optimal length is proportional to the square of MPW radius and is independent of laser amplitude. Typically, a THz radiation pulse at the order of a few tens of millijoule can be obtained from a joule level pump laser, corresponding to a peak efficiency over $6\%$. In addition, it is found that preplasma is detrimental to THz generation due to the distorted distribution of optical modes.

Our results directly assist the design of MPW THz sources that can reach unprecented conversion efficiencies among laser-plasma based approaches, and provide theoretical insights into the relevant microphysical processes.

\ack 
The authors acknowledge fruitful discussions with I Pusztai.
This work is supported by the Olle Engqvist Foundation and
the European Research Council (ERC-2014-CoG grant
647121). Simulations were performed on resources at
Chalmers Centre for Computational Science and Engineering (C3SE) provided by the Swedish National Infrastructure for Computing (SNIC).

\end{document}